%% file: main.tex
\documentclass[sigconf,nonacm,preprint,10pt]{acmart} 

\usepackage{amsmath,amsfonts}
\usepackage{bm}
\usepackage{graphicx}

\begin{document}

\title{Feasible but Not Safe: Constraint Violations and Report-Channel Attacks in Learned Cell-Free ISAC Association}

\thanks{\copyright 2026 Copyright held by the author(s). This is the author's version of the work. It is posted here for your personal use. Not for redistribution. The definitive Version of Record was published in ACM S3 '26, \url{https://doi.org/10.1145/3842433.3843753}.}

\author{Mehdi Zafari}
\affiliation{%
  \institution{University of California Irvine}
  \city{Irvine}
  \state{CA}
  \country{USA}
}

\author{Iman Mohammadi}
\affiliation{%
  \institution{University of California Irvine} 
  \city{Irvine}
  \state{CA}
  \country{USA}
}

\author{A. Lee Swindlehurst}
\affiliation{%
  \institution{University of California Irvine}
  \city{Irvine}
  \state{CA}
  \country{USA}
}

\input{sections/sec_abstract}

\maketitle

\input{sections/sec_intro}
\input{sections/sec_background}

\input{sections/sec_feasibility}

\input{sections/sec_safety}

\input{sections/sec_conc}

\bibliographystyle{ACM-Reference-Format}
\bibliography{refs}

\end{document}

%% file: sections/sec_abstract.tex
\begin{abstract}
Learning-based schedulers have been proposed to provide real-time user, target, and access point (AP) association in distributed cell-free integrated sensing and communication systems. In a typical approach, a graph neural network (GNN), trained on labels from a mixed-integer linear program, maps lightweight per-AP statistics to decisions on AP clustering, user and target scheduling, and mode selection in one forward pass. Such solutions assume that hard constraints, enforced only as soft training penalties, hold at inference, and that the self-reported statistics are truthful. Using our ASSENT algorithm as an example, we find that despite high $F_1$ scores, many solutions violate at least one hard constraint, demonstrating that high prediction accuracy does not ensure joint feasibility. Projecting the GNN output onto a feasible solution restores constraint satisfaction with low utility loss, even with a simple greedy repair procedure. We further show that feasibility alone does not guarantee robustness to false data injection attacks. A single malicious AP that reports false information cannot substantially increase its user associations, but can greatly increase the rate of infeasible solutions. The effect of such attacks depends on the type of information being falsified. Misreporting information that affects the objective can largely be mitigated through feasibility projection, whereas falsifying information that affects the constraints cannot. The latter can, however, be detected using a low-complexity cross-AP consistency check. These results show that learned ISAC schedulers should be evaluated using constraint-aware feasibility metrics in addition to conventional accuracy measures. 

\end{abstract}

\keywords{cell-free ISAC, graph neural networks, learning to optimize, feasibility, false data injection, 6G, wireless security}

%% file: sections/sec_intro.tex
\section{Introduction}
\label{sec:intro}


Cell-free architectures, introduced to manage interference and increase spectral efficiency for wireless communications~\cite{ngo2017cellfree}, are being considered for integrated sensing and communication (ISAC)~\cite{liu2022isac, demirhan2025cellfree}, where a large number of access points (APs) cooperate to serve users and to illuminate and sense targets.
A core control task is association: deciding, per resource block, which APs serve which users, which APs illuminate or receive echoes from which targets, and in which mode each AP operates, subject to radio frequency (RF) chain budgets, interference limits, and fronthaul constraints.
The association optimization task can be formulated as a mixed-integer linear program (MILP)~\cite{assent2025,abantoleon2025scheduling, memisoglu2024scheduling}, which is prohibitive to solve per slot at scale.
To overcome this computational bottleneck, we adopt a learning-to-optimize framework in which a graph neural network (GNN) learns from offline MILP solutions, allowing the system to bypass the exact solver using a single forward pass during inference~\cite{wangwong2024hetgnn, jiang2025graphlearning}.
The ASSENT scheduler~\cite{assent2025} is a representative example of such a solution, where each AP reports only low-dimensional statistics (average channel gains, inter-user correlations, sensing features) and a heterogeneous GNN maps them to the optimal association.

This design is lightweight, but its efficiency rests on two fragile assumptions.
First, hard constraints (e.g., RF budgets, half-duplex limits, etc.) act only as soft training penalties; at inference, the GNN thresholds continuous probabilities into discrete binary decisions on a per-variable basis, bypassing any explicit verification of joint feasibility.
Second, the central server (CS) cannot audit the statistics each AP reports, enabling malicious false data injection (FDI).
Thus, the inferred association decisions may lack both physical feasibility and operational trustworthiness.
Using the ASSENT approach in~\cite{assent2025}, we show these vulnerabilities are inextricably linked.
Rather than proposing a new architecture, we present the following empirical study:
\begin{itemize}
  \item The high-accuracy learned ASSENT ISAC scheduler produces infeasible plans for approximately $70\%$ to $74\%$ of its solutions. We explain why per-decision metrics obscure this and why constraint satisfaction is a necessary evaluation metric (Sec.~\ref{sec:feasibility}).
  \item Restoring feasibility is nearly free in terms of utility. A nearest-feasible projection, or even a greedy repair, changes the optimization objective by less than $0.1\%$ across the communication-sensing trade-off (Sec.~\ref{sec:feasibility}).
  \item The impact of a false data injection (FDI) attack depends fundamentally on the manipulated parameters. We demonstrate that projection mechanisms naturally neutralize attacks targeting the scheduling objective, whereas attacks falsifying physical constraints require a distributed cross-AP consistency check (Sec.~\ref{sec:safety}).\footnote{Source code, evaluation scripts, and replication data are publicly available at \url{https://github.com/LS-Wireless/ASSENT-Security-Analysis}.}
\end{itemize}



Although using soft-penalty constraints and post-hoc projection mechanisms to enforce rule compliance is standard practice in the machine learning optimization literature~\cite{donti2021dc3, liang2024homeomorphic}, our primary contribution is conducting an empirical audit of these techniques on a practical, distributed ISAC scheduler.
Importantly, we move beyond benign operational errors to establish a connection between a system's baseline feasibility and its susceptibility to targeted security threats.
Furthermore, while existing literature extensively covers traditional adversarial attacks against wireless models~\cite{manoj2021adversarial, le2025fggm, tao2021singlenode} and robust training defenses~\cite{blanchard2017byzantine, zhangzitnik2020gnnguard}, we investigate a fundamentally different threat vector.
Specifically, we expose the distinct runtime vulnerabilities that emerge when a central server must rely on unverified, self-reported channel and sensing statistics from potentially dishonest access points.

%% file: sections/sec_background.tex
\section{Background and Threat Model}
\label{sec:setup}


\textbf{Learned association.} We adopt the configuration assumed in ASSENT~\cite{assent2025}: $8$ APs ($4$ RF chains each), $10$ users, and $4$ targets.
Each AP reports local statistics, including communication gains, a user-to-user interference table, and sensing channel estimates.
A heterogeneous GNN processes these reports to output the AP mode $\tau$, AP-user association $x$, target scheduling $s$, and transmit/receive AP-target links $y^{\mathrm{tx}}, y^{\mathrm{rx}}$.
This network is trained to imitate MILP-optimal labels, including the hard constraints as soft penalties during training.

\textbf{Hard rules.} Formally, a generated association plan is strictly feasible if and only if it resides within the feasible set $\mathcal{F}$ defined by the MILP formulation.
Specifically, we evaluate the association for satisfying the following physical constraints: APs may serve users or illuminate targets only in transmit mode, and receive echoes only in receive mode; the sum of served users and transmit targets at an AP cannot exceed its number of RF chains; no user pair with a reported correlation exceeding $\rho_{\mathrm{th}}$ is co-served; each scheduled target is assigned at least one transmitting and one receiving AP; and conversely, if a target is not scheduled, no AP may illuminate or listen for its echo.\footnote{Specifically, we strictly evaluate constraints (7c) through (7k) from the ASSENT MILP~\cite{assent2025}. Optional capacity caps (7l) to (7n) are excluded.}
Validating this exact checker on $1000$ MILP-optimal plans, it correctly reports zero violations.

\textbf{Threat model.} The CS cannot directly audit per-AP reports.
A single malicious AP $a^\star$ can alter only its own reported statistics (gains, correlations, and sensing features) prior to processing.
The CS constructs the global plan based on these potentially falsified reports, and we measure the downstream impact on the true, uncorrupted channel.
We sweep the attacker position across $a^\star \in \{0,3,6\}$ since the resources an AP can acquire depend heavily on the network geometry.
To establish a worst-case baseline, we allow the attacker to scale its reported values up to $32\times$, acknowledging that while this demonstrates vulnerability, such unclipped extremes might be readily filtered as out-of-distribution anomalies in practical deployments.

%% file: sections/sec_feasibility.tex
\section{Good Scores, Broken Rules}
\label{sec:feasibility}


\textbf{High accuracy hides infeasibility.} ASSENT's per-decision $F_1$ is high (at least $0.84$ across all five outputs).
Yet, $73.5\%$ of the association results generated by the baseline Graph Attention Network (GATv2) violate at least one constraint ($69.2\%$ and $72.8\%$ for the alternative NNConv and TransformerConv graph layers, respectively).
The dominant violation is the RF-chain budget ($54\%$ to $58\%$ of plans, since nothing counts the number of AP RF chains at inference), followed by co-serving interfering users ($20\%$ to $26\%$) and mode errors ($18\%$ to $20\%$).

\begin{figure}[t]
  \centering
  \includegraphics[width=\linewidth]{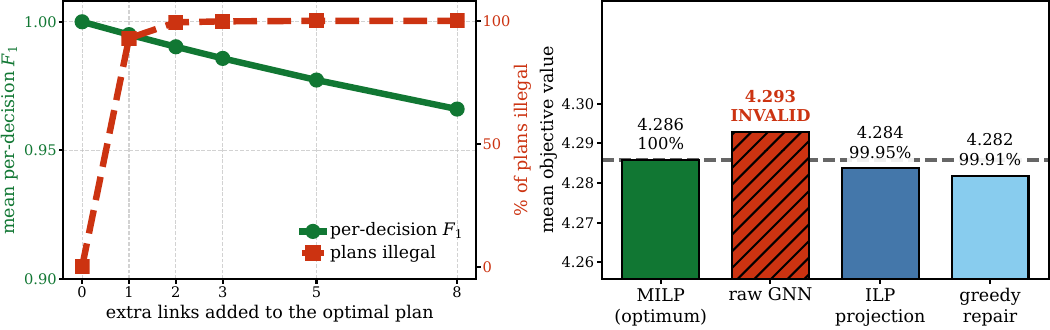}
  \Description{Two-panel figure evaluating scheduler feasibility. Left: adding a single excess link to optimal plans keeps F1 at 0.995 while 93 percent of plans become illegal. Right: bar chart showing MILP optimum, raw GNN, ILP projection, and greedy repair objective values, demonstrating that feasibility repairs stay within 0.1 percent of the optimum.}
  \caption{Feasibility of the learned scheduler (GATv2). Left: $F_1$ is blind to joint legality; adding one excess link keeps $F_1=0.995$ while $93\%$ of plans become illegal. Right: Exact ILP projection and greedy repair remain within $0.1\%$ of the MILP optimum; raw GNN scores are higher only due to illegal constraint violations.}
  \label{fig:feasibility}
\end{figure}

\textbf{Why $F_1$ is blind.} Standard classification metrics such as the $F_1$ score evaluate individual assignments independently, whereas system feasibility is a joint property of the global resource allocation.
From the optimal plans ($F_1{=}1$, all legal), adding one random excess link keeps $F_1$ at $0.995$ yet makes $93\%$ of plans illegal (Fig.~\ref{fig:feasibility}, left).
With exactly $D=156$ binary choices per plan in this setup, flipping one moves the score by only $1/D$ but can break legality outright; with $D$ in the hundreds, the metrics decouple.
A symptom appears in the baseline's performance results, where the raw GNN appears to beat the MILP optimum ($4.293$ vs.\ $4.286$).\footnote{Our raw GNN objective of $4.293$ differs slightly from the $4.311$ reported in~\cite{assent2025} because we evaluate on a fresh test seed to obtain paired confidence instances, holding the MILP baseline constant.}
This is not a real gain, since an illegal plan scores higher only by ignoring the rules it breaks, and the phantom gain vanishes once the plan is made legal.

\textbf{Restoring feasibility has negligible utility loss.} Treating the network outputs as confidences $p_v$, we formulate the $L_1$-nearest feasible plan as:
\begin{align}
&\max_{x,\tau,s,y^{\mathrm{tx}},y^{\mathrm{rx}}\in\{0,1\}} \ \sum_{v} (2p_v-1)\,v
\quad \nonumber \\
&\quad \quad \; \text{s.t.} \quad (x,\tau,s,y^{\mathrm{tx}},y^{\mathrm{rx}}) \ \in \ \mathcal{F},
\label{eq:proj}
\end{align}
which is an exact integer linear program (ILP) with $\mathcal{F}$ representing the feasible set defined in Sec.~\ref{sec:setup}.
This is legal by construction, so the headline is the utility cost, which is small and steady (Fig.~\ref{fig:feasibility}, right).
The projection retains $99.95\%$ of the true MILP scheduling objective for GATv2 ($99.98\%$ for the others), with the objective degradation averaging $0.06\%$ across the entire communication-sensing Pareto frontier (per-instance paired $95\%$ CI: $[0.04\%, 0.11\%]$).
The network's mistakes are small, i.e., an invalid plan results in only $1.17$ excess links on average.
Thus, a trivial greedy repair, which iteratively deletes the lowest-confidence offending choices to resolve capacity/clash violations and greedily adds the highest-confidence links to resolve target-coverage gaps, retains $99.91\%$ of the objective, performing within $0.03\%$ to $0.10\%$ of the exact ILP projection.
The diagnosis of the gap, rather than the repair itself, is the primary contribution.

%% file: sections/sec_safety.tex

\section{Legal Is Not Safe: Vulnerabilities to FDI}
\label{sec:safety}

\begin{figure}[t]
  \centering
  \includegraphics[width=\linewidth]{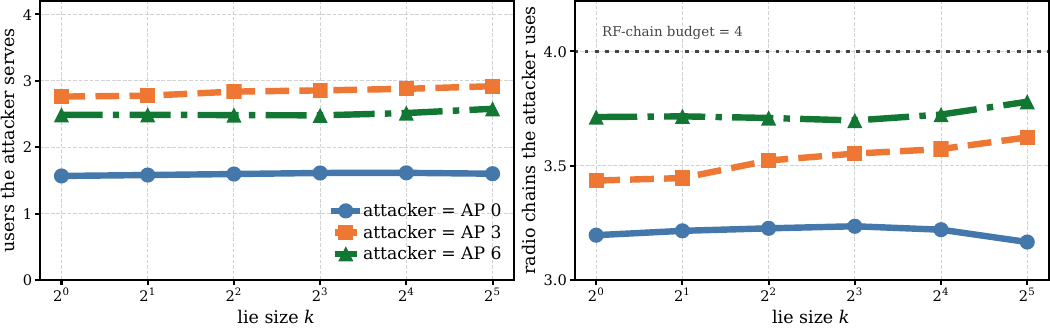}
  \Description{Two-panel line graph showing the effect of an AP inflating its reported signal. Left panel shows the number of served users barely increases. Right panel shows the attacker acquires minimal extra RF chains.}
  \caption{Signal FDI attack (steal variant) across three AP positions over the test set. Left: served users increase by at most $0.2$ even at $32\times$ inflation. Right: the attacker acquires minimal extra radios (budget $4$). Position variances stem from the symmetric deployment geometry relative to random users, not the GNN.}
  \label{fig:steal}
\end{figure}

A feasible association is not inherently secure: the nearest-feasible projection introduced in Sec.~\ref{sec:feasibility} operates after an FDI attack has already steered the GNN, meaning it cannot universally neutralize report-channel manipulations.

\textbf{Stealing is hard, breaking is easy.} Inflating the reported signal by attacker to acquire resources is ineffective (Fig.~\ref{fig:steal}): across all three positions, all GNN backbone layers, and scaling up to $32\times$, the attacker's served users increase by at most $0.2$ (e.g., AP0 from $1.57$ to $1.60$).
A greedy user-targeting heuristic performs no better, likely because gains are normalized relative to each AP's local range, making the assignment a joint choice over all nodes.
Conversely, sabotage is highly effective.
Under-reporting interference, falsely indicating to the CS that it is safe to co-serve mutually interfering users, pushes the proportion of infeasible associations from $73\%$ to $84\%$.
Falsifying sensing strength drives infeasibility to $96\%$ while enabling the attacker to monopolize target illumination links (averaging an increase from $0.7$ to $1.1$ allocated links), and two colluding APs reach $91\%$.
At $96\%$ illegal, almost every inference requires an operational fallback, creating a substantial reliability risk~\cite{assent2025}.

\begin{figure}[t]
  \centering
  \includegraphics[width=\linewidth]{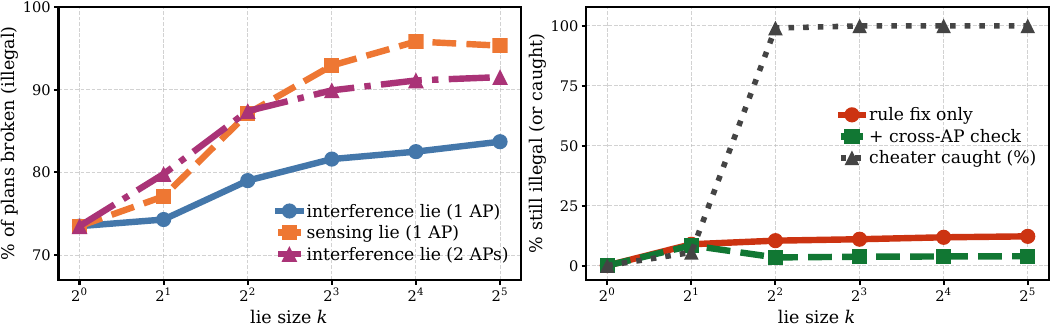}
  \Description{Two-panel line graph. Left shows percentage of illegal plans rising with attack size, peaking with the sensing lie. Right shows the cross-AP check catching the attacker near 100 percent of the time, dropping residual illegal plans to 4 percent.}
  \caption{Left: Report manipulations severely degrade system feasibility; sensing inflation and two-AP collusions are most disruptive. Right: Against the interference attack, projection leaves $12\%$ of plans physically infeasible; the cross-AP check detects the attacker ($99.5\%$ accuracy) and reduces the residual error to $4\%$.}
  \label{fig:defense}
\end{figure}

\textbf{What an attack corrupts dictates the defense.} Let $\Pi_{\mathcal F}(p)$ denote the projection~\eqref{eq:proj}.
If an FDI attack corrupts only objective coefficients (e.g., sensing gains), $\mathcal F$ remains unchanged and $\Pi_{\mathcal F}$ restores legality against the true physical rules: even the $96\%$ sensing attack is fully canceled ($0\%$ of corrected plans break a rule) because the constraints never read the falsified quantity.
If the attack corrupts a quantity that defines a constraint (e.g., interference values in the no-clash rule), $\mathcal F$ itself is falsified.
Consequently, $\Pi_{\mathcal F}$ enforces the wrong feasible set: about $12\%$ of the mathematically "legal" plans still violate a rule in physical reality, and the attacker (e.g., AP~0 under GATv2) is pushed back only from $1.5$ to $1.3$ users.
Thus, projection fully neutralizes an attack that cannot alter the feasible set, but only partially mitigates one that falsifies the constraints.

\textbf{A cross-AP consistency check.} Assuming a symmetric AP topology~\cite{assent2025}, AP reports contain inherent redundancy: every AP reports interference for the same user pairs, and a genuinely clashing pair appears correlated to many APs.
The server can flag any AP whose interference falls more than one median absolute deviation (MAD) below the group median, lifting the flagged values toward the group utilizing only reported data.
With a fixed threshold, this catches the attacker $\sim 99.5\%$ of the time ($99\%$ to $100\%$ range) with $\sim 1\%$ false alarms, cutting the residual broken share from $12\%$ to $4\%$ (Fig.~\ref{fig:defense}).
It effectively bounds the attacker: a large lie is almost always caught, while a small lie evades detection but barely perturbs the system.

%% file: sections/sec_conc.tex
\section{Discussion and Conclusion}
\label{sec:conclusion}

\textbf{Conclusion.} Evaluating learned cell-free ISAC schedulers requires constraint-aware metrics, as high per-decision accuracy often masks severe joint infeasibility. While post-hoc projection efficiently restores physical feasibility with near-zero utility loss, enforcing legality does not inherently guarantee security. Crucially, the effectiveness of projection against false data injection (FDI) depends entirely on the attacker's target: lies that corrupt the scheduling objective are neutralized, whereas lies that falsify the underlying physical constraints weaponize the repair mechanism itself.

\textbf{Scope and Limitations.} While these findings are consistent across multiple GNN backbones, our empirical evaluation is bounded by the topology of the released system ($8$ APs, $10$ users, $4$ targets). Evaluating larger networks requires computationally heavy exact solvers to generate new training labels, so verifying scalability remains future work.

\textbf{Toward Feasible and Trustworthy Association.} Resolving these vulnerabilities requires moving beyond post-hoc repairs. At the model level, constraint-aware decoding~\cite{donti2021dc3, tang2024l2o} could force the GNN to emit inherently feasible plans, eliminating baseline violations at their source. At the system level, implementing robust, multi-agent verification checks~\cite{blanchard2017byzantine, zhangzitnik2020gnnguard} is necessary to harden the reporting channel against sophisticated or colluding attackers. Ultimately, these MAC-layer insights motivate a secure PHY-MAC co-design, where a robust, feasibility-aware associator reliably feeds legal, trustworthy schedules to a coordinated PHY-layer resource allocator~\cite{zafari2025coordinated}.